
\magnification=1200
\overfullrule=0pt
\baselineskip=15pt
\font\mfst=cmr9
\font\mfs=cmr9 scaled \magstep 1
\font\rma=cmr9 scaled \magstep 2
\font\rmm=cmr9 scaled \magstep 3
\def\ueber#1#2{{\setbox0=\hbox{$#1$}%
  \setbox1=\hbox to\wd0{\hss$\scriptscriptstyle #2$\hss}%
  \offinterlineskip
  \vbox{\box1\kern0.4mm\box0}}{}}

\def\hn{$\;{\rm h}^{-1}$}

\def\R{\rm I\kern-.18em R}
\def\etal{{\it et al. }}
\topskip 6 true cm
\pageno=0

\centerline{\rmm TESTING HIGHER--ORDER LAGRANGIAN}
\smallskip
\centerline{\rmm PERTURBATION THEORY}
\smallskip
\centerline{\rmm AGAINST NUMERICAL SIMULATIONS --}
\bigskip
\centerline{\rmm 2. HIERARCHICAL MODELS}
\bigskip\bigskip
\centerline{\rmm by}
\bigskip\bigskip
\centerline{\rmm A.L. Melott$^{1}$, T. Buchert$^{2}$, A.G.
Wei{\ss}$^{2}$}
\vskip 1.5 true cm
\centerline{\rma $^{1}$Department of Physics and Astronomy}
\smallskip
\centerline{\rma University of Kansas}
\smallskip
\centerline{\rma Lawrence, Kansas 66045}
\smallskip
\centerline{\rma U. S. A.}
\bigskip
\centerline{\rma $^{2}$Max-Planck-Institut f{\"u}r Astrophysik}
\smallskip
\centerline{\rma Postfach 1523}
\smallskip
\centerline{\rma 85740 Garching, Munich}
\smallskip
\centerline{\rma F. R. G.}

\vskip 1 true cm
\centerline{\mfs submitted to {\it Astronomy and Astrophysics},
Main Journal}

\vfill\eject

\topskip= 2 true cm
\centerline{\rmm Testing higher--order Lagrangian perturbation theory}
\smallskip

\centerline{\rmm against numerical simulations --}

\medskip

\centerline{\rmm 2. Hierarchical models}
\bigskip\medskip
\centerline{\rma by}
\bigskip\medskip
\centerline{\rma A.L. Melott, T. Buchert, A.G. Wei{\ss}}
\vskip 0.8 true cm
\noindent
{\mfst
{\narrower
\baselineskip=12pt
{\mfs Summary:}
We present results showing an improvement of the accuracy of
perturbation theory as applied to cosmological structure formation for
a useful range of scales.
The Lagrangian theory of gravitational instability of
Friedmann--Lema\^\i tre cosmogonies
investigated and solved
up to the third order in the series of papers by
Buchert (1989, 1992, 1993), Buchert \& Ehlers (1993), Buchert (1994),
Ehlers \& Buchert (1994),
is compared with numerical simulations.

\noindent
In this paper we study the dynamics of hierarchical models
as a second step. In the first step (Buchert, Melott and Wei{\ss} 1994)
we analyzed
the performance of the Lagrangian schemes for pancake models, i.e.,
models which initially have a truncated power spectrum.
We here explore whether the results found for pancake models carry over
to hierarchical models which are evolved deeply into the non--linear
regime.
We smooth the initial data by using a variety of filter types and
filter scales in order to determine the optimal performance of the
analytical models, as has been done for
the ``Zel'dovich--approximation'' -- hereafter TZA -- (as a subclass of
the irrotational Lagrangian first-order solution) in
previous work (Melott \etal 1994a).
We study
cross--correlation statistics employed in previous work for power--law
spectra having indices in the range $(-3,+1)$.

\noindent
We find that for spectra with negative power--index
the second--order scheme performs considerably better than TZA
in terms of statistics which probe the dynamics,
and slightly better in terms of low--order statistics like
the power--spectrum.
In cases with much small--scale power the gain from the higher--order
schemes is small, but still measurable.
However, in contrast to the results found for pancake models, where the
higher--order schemes get worse than TZA at late non--linear
stages and on small scales, we here find that the second--order
model is as robust as TZA, retaining the improvement at later
stages and on smaller scales. In view of these results we expect that
the second--order truncated Lagrangian model is especially
useful for the modelling of standard dark matter models such as
Hot--, Cold--, and Mixed--Dark--Matter.

}}

\vfill\eject
\topskip= 0 true cm
\baselineskip=14.5pt

\noindent{\rmm 1. Introduction}
\bigskip\bigskip
\noindent
The non--linear modelling of inhomogeneities initiated by
small--amplitude fluctuations at the time of decoupling of matter and
radiation has been, for a long time, the privilege of numerical N--body
simulations. Recently, the interest in the question has grown how well
this modelling
can be achieved by analytical or semi--numerical methods. Most
of those efforts concentrate on Lagrangian methods introduced in a
pioneering work on the Zel'dovich (1970, 1973) approximation (hereafter
ZA).
Here, we pursue a systematic ana\-ly\-sis of analytic solutions obtained
in the framework of the Lagrangian instability picture investigated
by one of us (Buchert 1989, 1992). Lagrangian perturbation solutions
are available
up to the third order (Buchert 1994).
In this paper we restrict the presentation of results
to the first and second--order schemes (Buchert \& Ehlers 1993, Buchert
1993a).

\medskip
Parallel efforts in this direction, though less systematic, have been
undertaken by Moutarde \etal (1991), Bouchet \etal (1992), Gramann
(1993), and Lachi\`eze--Rey (1993a,b).
Other approximations are also built in a similar spirit, for instance
the ``adhesion model'' (Gurbatov \etal 1989)
introduces an artificial viscosity term into Zel'dovich's formalism,
the ``frozen--flow--approximation'' (Matarrese \etal 1992) assumes
constancy of the velocity field, the
``frozen--potential--approximation'' (Brainerd \etal 1994, Bagla \&
Padmanabhan 1994) assumes
constancy of the gravitational potential. All of these approaches
are able to some extent to mimic the results of
N-body simulations
(for the first see Kofman \etal 1992 and Melott \etal 1994c, for the
second, Melott \etal 1994b, for the third see either reference).
All of them attempt to
repair a major shortcoming of ZA which predicts
uncomfortably thick pancakes after shell--crossing: In the
``adhesion model'' a viscosity term mimics a `sticking' of pancakes.
In the ``frozen--flow--approximation''
only the first--generation pancakes are reached and
only asymptotically. The ``frozen--potential--approximation'' (as a
semi--numerical model) counteracts the
thickening of pancakes by modelling secondary shell--crossings,
a feature
which is observed in numerical simulations as well as in the
higher--order Lagrangian schemes analyzed here (compare Buchert \&
Ehlers 1993), and can therefore be attributed to part of the
``true'' effect of self-gravity. However, it is almost as time consuming
as a
full (Particle--Mesh) N--body simulation.
Other work to be mentioned in this context are the articles by
Matarrese \etal (1993), Kasai (1993), Croudace \etal (1994),
Bertschinger \& Jain (1994), Berschinger \& Hamilton (1994), Kofman \&
Pogosyan (1994), Salopek \etal (1994),
dealing with the corresponding equations of General Relativity,
or their Newtonian limit, respectively, and Giavalisco \etal (1993),
Munshi \& Starobinsky (1994), Munshi \etal (1994).
Matarrese \etal (1994) presented a general relativistic second--order
Lagrangian solution.

\medskip
There are two major problems with most of the tests (if there were any
tests)
of various schemes discussed
here: (1) Some of them were tested by different techniques for
accuracy, and against
different simulations. This makes it impossible to compare their
accuracy. (2)
Often CDM spectra were used (e.g., Bagla \&
Padmanabhan 1994). Since the spectral index of CDM is negative on
scales going
non--linear (more so with biasing), it is not hard to get it more or
less correct with ZA or ZA--based approximations.
\medskip
Coles, Melott and Shandarin (1993) (hereafter CMS) and Melott, Pellman
and Shandarin (1994a) (hereafter MPS) conducted a series
of tests of analytical
approximations by cross--correlating them with N--body
simulations. They tested the
Eulerian linear
perturbation solution and ZA, i.e.,
the Lagrangian linear solution. They found that the latter is
considerably more successful than the former. Considerable further
improvement is made if ZA is applied
not to the full power spectrum, but
only to a {\it truncated} body of the spectrum (TZA). The truncation
removes unwanted non--linearities, which are evolved beyond the range of
applicability of the model.
The shape of high--frequency filters imposed on the power--spectrum was
found to be important (MPS).
Most successful in respect of improvement of the cross--correlation
statistics was the use of Gaussian filters,
but if one wishes to work backwards
from evolved to initial state,
a sharp truncation (step function) in $k$ is
preferable (Melott 1993).

\medskip
Most recent work has applied these tests to compare
TZA with the ``adhesion model'' and the ``frozen--flow--approximation''.
The result was that those procedures provide reasonable
statistical toy models, however, they are both {\it dynamically} less
accurate than TZA as
demonstrated by the comparison of their cross--correlation statistics
in the fully developed non--linear regime
(Melott \etal 1994b,c). The comparison of higher moments --
skewness and kurtosis -- of the density contrast in the weakly
non--linear regime (Bernardeau
1994, Munshi \& Starobinsky 1994, Munshi \etal 1994)
has also demonstrated that
the ``frozen--flow--'' and ``frozen--potential--'' approximations
are dynamically less accurate than ZA. However, the results of the
latter tests cannot be extrapolated to late non--linear stages,
where also ZA has to be replaced by TZA.

\medskip
Because of the success of TZA it is worthwhile to study higher--order
corrections to ZA as a subclass of the first--order Lagrangian
perturbation solution (Buchert 1992) and also truncate the initial data
in these models to see whether the performance of TZA can
be further improved.
We here continue our study on the Lagrangian
perturbation schemes begun by Buchert, Melott and Wei{\ss} (1994a),
henceforth BMW.
In that paper we concentrated on pancake models, i.e., models which do
not involve small--scale power in the initial conditions. We did
this as a first step in
understanding the limitations of higher--order
corrections to Zel'dovich's mapping. Also, pancake models can be
understood as generic archetype of hierarchical models which involve
pancaking on all spatial scales (Kofman \etal 1992). In
a second step,
we here analyze (in the spirit of MPS) various power--spectra which are
evolved deeply into the
non--linear regime in order to probe the performance of the Lagrangian
approximations in the case of hierarchical models.
Although, it is not expected that a perturbation approach
can be used to model highly non--linear stages, we here demonstrate
that one can obtain good agreement with the N--body simulation
by smoothing the initial data before evolving them.

\medskip
BMW found that the second--order approximation provides in pancake
models a definite and useful
improvement upon the first--order approximation (ZA).
We explore now whether this improvement carries over to hierarchical
models.

\vfill\eject

\noindent{\rmm 2. Numerical realization, Lagrangian perturbation
schemes}
\smallskip
\centerline{\rmm and cross--correlation statistics}
\bigskip\medskip
\noindent
{\rma 2.1. Numerical realization}
\bigskip\noindent
We specify initial data in terms of the power--spectrum
${\cal P}(k)$
(as a function of comoving wavenumber $k=\vert \vec k \vert$)
of Gaussian density perturbations of the form:
$$
{\cal P} (k)=
<\vert \delta_{\vec k} \vert^2 > \propto \vert \vec k
\vert^n\;\;,\eqno(1)
$$
where $\delta_{\vec k}$ is the discrete spatial Fourier transform of the
density
contrast $\delta$, and $<...>$ denotes ensemble average over the entire
distribution, assumed to be equivalent to the spatial average.
We consider the range of indices $-3 \le n \le +1$; for all spectra
we use the same set of random phases, so that the overall structure is
similar in all realizations.

\medskip
We emphasize that we have to give initial data early in order to
fairly estimate the effect of the higher--order schemes and to
guarantee an objective modelling of the collapse
of first objects in the model (see the discussion in BMW).
For the spectra with indices $n=-3,-1$ and $n=+1$ we give initial data
at the r.m.s.
density contrast of $\sigma_i (k_{N_y})= .05$), where $k_{N_y}$ is the
Nyquist frequency. For the other two indices we have taken a larger
amplitude ($\sigma_i (k_{N_y})= 0.2$), which is about that amplitude
people normally use. This we did in order to see, whether there
is any detectable effect. Indeed we found an effect which will be
reported in Section 3 (see also Melott {\it et al.} 1994c).

In order to let the concrete normalization of the spectra open,
we define the evolution stages of the realizations in terms of the
non--linearity scale $k_{nl}(t)$:
$$
a^2(t) \; \int_0^{k_{nl}(t)} \; d^3 k \; {\cal P}(k)
\;=\;1\;\;,\eqno(2)
$$
where $k_{nl}(t)$ is decreasing with time as successively larger scales
enter the non--linear regime; $a(t)$ is the scale factor of the
homogeneous background ($a(t_i):=1$).
The evolution of inhomogeneities is modelled in a flat
Einstein--de Sitter background ($a(t)=({t \over t_i})^{2/3}$).
We shall study two stages corresponding to $k_{nl}=8 k_f$ and
$k_{nl} = 4 k_f$, where
$k_f$ is the fundamental mode of the simulation box.
Thus, our models studied here have been evolved for expansion factors of
240 to 5100, depending on spectral index and $k_{nl}$.
We shall present all statistical results for both stages.
\medskip
We evolve the fields with an enhanced PM (Particle--Mesh) method
(Melott 1986) using $128^3$ particles
each on a comoving $128^3$ mesh with periodic boundary conditions.
This method makes the code resolution--equivalent to a traditional
PM code with $128^3$ particles on a $256^3$ mesh (see Park 1990, 1991,
and Weinberg \etal 1994). We then bin the data into a $64^3$ mesh
using the CIC scheme. Due to our controls (the same initial phases in
all
simulations) it is not necessary to do an ensemble of simulations with
each
power spectrum. One of each spectral type is sufficient to uncover
systematic
effects due to changing approximations.
For more details see CMS.

\vfill\eject
\noindent
{\rma 2.2. Lagrangian perturbation schemes}
\bigskip\noindent
For the Lagrangian perturbation schemes up to the third order and
their realization see Buchert (1994) and BMW.
However, we here restrict the presentation of results
to the first-- and second--order schemes, the second--order scheme is
expected to contain the major effects on large scales according
to the results of BMW for the present purpose. But, we also made
a consistency test by going to the third order.

\medskip
Denoting comoving Eulerian coordinates by $\vec q$ and Lagrangian
coordinates by $\vec X$,
the field of trajectories $\vec q = \vec F (\vec
X,t)$ up to the second order reads:
$$
\vec F = \vec X \;+\;
q_1 (a) \; \nabla_0 {\cal S}^{(1)} (\vec X) \;+\; q_{2}
(a) \; \nabla_0 {\cal S}^{(2)} (\vec X)
\;\;,\eqno(3)
$$
with the time--dependent coefficients expressed in terms of the
expansion function $a(t) = ({t \over t_i})^{2/3}$:
$$
\eqalignno{
q_1 &= \left({3 \over 2}\right) (a - 1) \;\;\;, &(3a) \cr
q_{2} &= \left({3 \over 2}\right)^2
(-{3 \over 14} a^2 + {3 \over 5} a - {1 \over 2}
+ {4 \over 35} a^{-{3 \over 2}}) \;\;\;. &(3b) \cr}
$$
The perturbation potentials have to be constructed by
solving iteratively the $2$ Poisson equations:
$$
\eqalignno{
&\Delta_0 {\cal S}^{(1)} = I({\cal S}_{,i,k})t_i  \;\;\;,&(3c)
\cr
&\Delta_0 {\cal S}^{(2)} = 2 II({\cal S}^{(1)}_{,i,k})
\;\;\;, &(3d) \cr}
$$
where $I$ and $II$ denote the first and second principal
scalar invariants of the tensor gradient
$({\cal S}^{(1)}_{,i,k})$:
$$
I({\cal S}^{(1)}_{,i,k}) = tr({\cal S}^{(1)}_{,i,k}) =
\Delta_0 {\cal S}^{(1)} \;\;\;, \eqno(3e)
$$
$$
II({\cal S}^{(1)}_{,i,k}) = {1 \over 2}
\lbrack(tr({\cal S}^{(1)}_{,i,k}))^2 -
tr(({\cal S}^{(1)}_{,i,k})^2)\rbrack \;\;\;.\eqno(3f)
$$
The general set of initial data is
restricted according to the assumption of parallelism of the
peculiar--velocity field and the peculiar--acceleration field
at the initial time $t_i$ (see Buchert 1994
and ref. therein for a discussion of this
restriction). Therefore, the initial fluctuation field
can be specified to the peculiar--velocity potential $\cal S$ alone.
We can set ${\cal S}^{(1)}={\cal S}t_i$ (which is the unique solution
of the first Poisson equation (3c) for periodic initial data, see
Buchert 1992, Ehlers \& Buchert 1994). Doing this, the flow--field (3)
reduces to ZA if restricted to the first order.
%
%
\bigskip\smallskip
We realize the solution by first solving Poisson's equation for $\cal S$
via FFT (Fast Fourier Transform) from
the initial density contrast $\delta$ generated as initial condition
for the numerical simulation.
In an Einstein--de Sitter model we have:
$$
\Delta_0 {\cal S} = - {2 \over 3 t_i} \delta \;\;.\eqno(4)
$$
We then calculate the second principal invariant $II$ directly from
$\cal S$ and solve the second Poisson equation (3d) using FFT.
The density in the
analytical models is calculated by
collecting trajectories of the Lagrangian perturbation
solutions at the different orders into a $64^3$ pixel grid
with the same method (CIC binning) as in the
N--body simulation.

The CPU times on a CRAY YMP are for the first--order scheme $25$
seconds, and for the second--order scheme $60$ seconds;
the corresponding CPU times on a CONVEX C220 are $2$ and $5$ minutes.
Thus,
even the second--order
scheme is competitive with one step in a corresponding
PM--type N--body simulation.

\vfill\eject
\noindent
{\rma 2.3. Cross--correlation statistics}
\bigskip\noindent
For the presentation of the cross--correlation statistics we
refer the reader to CMS, MPS and BMW for details. We use four
statistics.
Firstly, the usual cross--correlation coefficient $S$ to compare
the resulting density fields:
$$
S := {<(\delta_1 \delta_2)> \over \sigma_1 \sigma_2} \;\;, \eqno(5)
$$
where $\delta_{\ell}, \ell=1,2$  represent the density contrasts in
the analytical and the numerical approximations, respectively,
$\sigma_{\ell} = \sqrt{<\delta_{\ell}^2>-<\delta_{\ell}>^2}$
is the standard deviation in a Gaussian random field;
averages $<...>$ are taken over the entire distribution. We believe this
is the
most important statistical test, because it measures whether the
approximation
is moving mass to the right place, with an emphasis on dense regions. We
allow
for small errors by presenting $S$ for the two density arrays smoothed
at a
variety of smoothing lengths.

Secondly, the
power spectrum (eq. (1)) of the evolved N--body model and the
analytic approximations were calculated. Thirdly, the mass density
distribution for both of them
are also shown. Both of these are widely displayed diagnostics in
cosmology.
\medskip
MPS found that the primary reason for the superiority of one window
function
over another (equations (6) below) lay not in the final power spectral
amplitudes
but in the phase angle accuracy. We therefore follow MPS and will
display
$<\cos\theta>_k$, where $\theta=\phi_1-\phi_2$ is the difference in the
phase
angle of the Fourier coefficients of mass density between the
approximation and the
simulation. The averaging is over all coefficients with the same
magnitude of wave vector, as in the power--spectrum.


\vfill\eject

\noindent{\rmm 3. Filtering the initial data}
\bigskip\medskip
\noindent
For the analytical realizations we filter the power--spectrum
on various scales $k_c$ using a set of ``trial windows''.
The use of Gaussian windows
proved to yield the best cross--correlation with the N--body simulations
in previous work on TZA (see MPS). We here again test three main types,
sharp $k-$truncation (a), Gaussian windows (b),
and tophat--smoothing (c).
The filters (a) and (b) are described in Fourier space, whereas the
filter
(c) is applied in real space (but the computing for (c) is done in
Fourier
space). The initial
data are modified to $\delta^*_{\vec k} = W  \delta_{\vec k}$,
where the different filters $W$ are:
$$
W_{tr}(k;k_{tr})=\cases{1,&$k\leq k_{tr}$\cr 0,&$k> k_{tr}$\cr}
\eqno(6a)
$$
$$
W_G (k; k_G) = e^{-k^2 / 2k_G^2}  \eqno(6b)
$$
$$
W_{th} (k; R_{th})=\biggl({\sin(R_{th}k)\over
(R_{th}k)^3}-{\cos(R_{th}k)\over
(R_{th} k)^2}\biggr)   \;\;\;.\eqno(6c)
$$
We explored a wide range of values for the scale of each possible filter
to find the optimum scale $k_{opt}$ for each type, as done for
first--order (ZA) by MPS.
\medskip
In Figs.1 we display the cross--correlation coefficient (eq. (5)) as
obtained
from a comparison of the truncated analytical schemes with the raw data
of the numerical simulations against the truncation scale
for all spectra and for all filter types.
This shows how the
optimal truncation (defined as the scale with the largest
cross--correlation coefficient) is varying with scale.
In the cases with negative power index
we notice that the variation in $S$ with scale is
much smaller towards higher $k$ than in the cases with positive power
index. The variation
effect is very weak in cases with small--scale power index less than
$n=-1$.
\medskip
While type (c) gives much better agreement with the N--body results
than simple $k-$trun\-cation (a), we find as in MPS
that the best filter type is always (b). Although the values for
$S$ for the best truncation are numerically similar for type (b) and
(c), the truncation scale for the best agreement is always larger for
type (c), thus erasing more non--linear information.
We therefore concentrate our further analysis on type (b) and recommend
generic use of Gaussian filtering.
\medskip
In Table 1
we display the values of $S$ for the optimal truncation
scales $k_{opt}$ for the different power--spectra and Lagrangian
schemes.
In Fig.2 we illustrate this table by plotting $S$ versus the power
index $n$.
We see an almost constant (but small) improvement of the
second--order scheme
over first--order for positive indices. As soon as the index drops
negative we find increasing improvement with decreasing power.
Fig.2 also shows that we obtained slightly different performance for the
models
with spectral index $n=0$ and $n=-2$. These models have been started
at a later stage (compare 2.1).
We think that this could be an indication for
inaccuracies of the numerical evolution code for low--amplitude
initial conditions, which we want to test in future work (see also
Melott {\it
et al.} 1994c).
\medskip
We have used discrete steps in $k$ to find the optimal truncation
lengths.
If we express the optimal scales in terms of the non--linearity scale,
e.g., $k_{nl}=8 k_f$, we find that the
arithmetic mean (used here as a spectrum independent estimate) lies for
the first--order scheme at $1.45 k_{nl}$, for second--order at $1.2
k_{nl}$, i.e., at a slightly larger scale. This difference almost
vanishes at the
later stage ($k_{nl}=4k_f$), where the arithmetic means are
$1.5 k_{nl}$ for first--order and $1.4 k_{nl}$ for second--order.
Still, within the
uncertainty bounds involved in the determination of the optimal
scale we can recommend generic use of a $k-$scale close to, but larger
than the non--linearity scale as suggested by MPS.
In real space this means that the comoving truncation scale $\lambda_T
< \lambda_{nl}$. It is interesting that, although the second--order
scheme requires slightly larger truncation scales, the dependence of
that scale on the spectrum is weaker than in the first--order scheme.
We emphasize that for negative--sloped spectra the truncation scale
$\lambda_T$
is especially small; in view of Fig.1 a truncation of models with
negative spectral slope is not as important as in the other models.
For spectra $n\leq -2$ the differences in $S$ on the small--scale end
are negligible. These models perform well even without truncation
as will become evident from the Figs.4 below.

\bigskip\noindent
We proceed by statistically analyzing the optimally truncated models
only, with $k_{opt}$ according to Table 1.
The results are presented and discussed in the next section.

\vfill\eject

\noindent{\rmm 4. Discussion of the results}
\bigskip\medskip
\noindent
We first concentrate on the first stage corresponding to $k_{nl}=8k_f$.

Fig.3 presents the cross--correlation coefficient
with the N--body
simulation as a function of scale for the optimally truncated and
untruncated
first-- and second--order schemes together with two versions of the
(still widely used) Eulerian linear approximation: (1) linear density
contrast, and (2) (following CMS): ``chopped'' linear density
contrast, i.e.: $1+\delta_{chopped}=\alpha (1+\delta)$, if $\delta > -1$
and $0$ otherwise, where $\alpha$ is the normalization constant keeping
the total mass the same.
We have chosen the spectra with indices $n=-1$ and
$n=+1$ for the illustration of the differences among the different
approximations.
This comparison again demonstrates that the Lagrangian schemes (even in
their untruncated form) perform
substantially better than the Eulerian solution with or without
``chopping''. In comparison with the following figures this
demonstration should teach that the Eulerian approach is almost useless
to describe even slightly non--linearly evolved density fields.
\medskip
In Figs.4 we display slices of the density contrast field as predicted
by the N--body simulations and the two optimally truncated
Lagrangian perturbation schemes.
We also display the evolved fields for the
untruncated perturbation schemes to illustrate the effect of
truncation, and for the ``chopped'' Eulerian first--order model.
The structures for different power--spectra bear family resemblance due
to our phase controls.

Generally, we appreciate a reasonable correspondence of the numerical
density fields with those of the truncated Lagrangian schemes.
Although, in the cases with small amount of power on small scales
(especially $n=-2,-3$) the pictures for the {\it untruncated}
models
might suggest better coincidence with the numerical density fields,
the cross--correlation statistics with the raw numerical data (shown in
Figs.1) prove that the truncated models agree slightly better.
However, in these models
the truncation plays a peculiar role, since
the cross--correlation coefficient only varies weakly as a function
of scale by going to smaller scales. The small numerical differences
in the coefficients can be seen in Figs.1.
Already small differences in the N--body computing can yield
different ``optimal'' truncation scales. Within this uncertainty, it is
a matter of standards we want to apply, whether we truncate the models
$n=-2,3$ or not.
But, clearly for models with spectra $n>-2$, truncation of the initial
data is indispensible.

The contours of the optimally truncated first-- and
second--order models are quite similar; visually manifest differences
only occur for the cases $n=-2$ and $n=-3$. Note that this is also
due to
the presentation in terms of grey--scale plots which are truncated
at small density levels (density contrast $5$), where both
schemes predict similar contours. The higher cross--correlation
coefficient of the second--order scheme reported below is the result
of sharper and higher density peaks at second order, visible
at higher density levels.

\medskip
In Figs.5 we show the r.m.s. density contrast of the N--body
simulations together with that of the analytical schemes
as a function of
scale. In the following figures we shall use the r.m.s. density
contrast of the N--body simulations as a scale parameter.
\medskip
Figs.6 show the most important result of this paper: the
cross--correlation coefficient for different smoothing scales (smoothed
with a Gaussian window) for different power--spectra and the two
truncated
Lagrangian schemes. We infer that the second--order approximation
shows higher cross--correlation with the N--body density field for
all spectra and for all scales. However, at this stage the improvement
is small in cases with non--negative power index, but it grows
substantially
with decreasing small--scale power. This result can already be
seen in Fig.2. In general, we find that
for negative power indices the agreement with the N--body field is
very good, especially on scales which are relevant for modelling
large--scale structure (i.e., above the scale of galaxy clusters),
bearing in mind that both the numerical
simulations and the analytical schemes are assumed to be
approximations to the unknown exact solution of the problem.
On scales far below the non--linearity scale,
where the r.m.s. density contrast of the N--body simulation
assumes the value $2$, the
cross--correlation coefficient generally lies close to $S=0.8$ in
the first--order scheme, while the second--order scheme is closer to
$S=0.9$ for negative--sloped power--spectra.

\medskip
While the cross--correlation test yields information on the dynamical
capabilities of the models, the following statistics show whether
they are useful in reproducing statistical properties of the density
field.

In Figs.7 we display the power--spectra as calculated from the
final distributions. We confirm the results by MPS who found
a weak representation of the power--spectrum on scales below the
non--linearity scale in TZA for positive power--spectra slopes.
Here, the second--order result does not improve upon
first--order.
Again, for negative slopes, the agreement is reasonably good for
both approximation schemes also below the non--linearity scale. In
summary, for much small--scale power the power--spectrum is weakly
represented below the non--linearity scale and especially below the
comoving truncation scale $\lambda_T$.
This is evident, since a truncation of the spectrum removes power which
cannot be compensated by non--linear evolution.
\medskip
The power--spectrum provides low--order statistical
information. Figs.8 and 9 present
statistics which probe higher--order
information of the density distribution. We find that
phase--information is represented much better in the
second--order
scheme in the cases $n=-2-3$, while in all other cases the performance
of second--order is moderately better than the first--order model.
Also here we see that the positive second--order
effect is counteracted by a large amount of small--scale power
at this stage which erases phase--information present in the
weakly non--linear regime.
\medskip
The density distribution functions (Figs.9)
show excellent agreement with the N--body result
for all cases with negative power index; for $n=0,+1$
the number of high--density cells is underestimated and the number of
low--density cells is overestimated by both schemes. The
improvement due to the second--order correction is still measurable.
\medskip
For the sake of showing consistency of our results, we made two
further studies: (1) we evolved the fields to a later non--linear
stage, where the non--linearity wavenumber has dropped to half the value
of the stage analyzed above; (2) we also calculated the statistics for
the third--order model (see BMW).

In Figs.10--14 the statistics corresponding to Figs.5--9 are
displayed for the later stage with $k_{nl}=4 k_f$.

One of the striking results obtained by MPS concerning the
cross--correlation
analysis is that TZA has proved to be very robust, i.e., the good
performance of this model remains stable by going to later stages, or to
smaller scales, respectively. In the context of the Lagrangian
perturbation solutions, BMW attributed this property to the fact
that TZA is the principal first part of a perturbation sequence, while
higher--order corrections display a ``blow--up'' phenomenon
at and after the stage when the perturbation theory breaks down.
Interestingly, we here find that
the improvement found for the second--order Lagrangian scheme
is also robust, i.e., at later stages and on small scales
the {\it truncated} second--order model does not become worse than
TZA as observed for pancake models by BMW.
This feature can be traced back to the optimal filtering employed here
which compensates the ``blow--up'' of the higher--order coefficients
by the choice of a different optimal scale for each order.
Moreover, the improvement of second-- upon first--order is even
larger at the later stage, especially with respect to the
cross--correlation coefficients, the phase--errors, and the density
distribution functions. In particular, also the spectra with positive
power--index show improvement, especially on small scales.
\medskip
Since from previous comparisons with other approximations we know that
it is hard
to improve on the dynamical performance of TZA, we consider this as a
definite and useful success of the second--order model.
It is remarkable that, although the second--order scheme
needs truncation at a slightly smaller frequency than first--order, thus
erasing more information initially (compare Fig.1), the non--linear
evolution overcompensates that.

\medskip
We could not confirm the same property of `robustness'
for the third--order scheme. In BMW we already noted that
differences between second-- and third--order are smaller than
the difference
found for second-- upon first--order, a result which is expected for
a perturbation approach which converges to the exact solution (if it
does). However, the break--down phenomenon at and after the perturbation
schemes are evolved beyond the range of validity after shell--crossing,
i.e. the behavior observed in pancake models by BMW, could be
compensated by optimized filtering of initial data for second--order
perturbations. This seems not to work for the third--order scheme.
Here, we have to stress that the numerical realization of the
third--order scheme by iteratively solving Poisson equations should be
very accurate in order to draw definite conclusions, especially for
models with much small--scale power. We think that much more numerical
effort has to be added in order to treat the third--order effect
reliably. Remember that in the third--order model $7$ Poisson
equations have to be solved, $4$ of them based on solutions of
the second--order Poisson equation, while in the second--order model
only $2$ Poisson equations have to be solved (directly from the
initial data).
Buchert \etal (1994b) therefore pursue a different way
to overcome numerical problems by solving Poisson equations
{\it analytically}
up to the third order.
With this method one can obtain
analytical expressions for all $7$ perturbation potentials,
and realize the Lagrangian scheme
for random fields with up to about $100$ Fourier modes at
reasonable CPU speed.
First results of this comparison
are discussed by Buchert (1993b).

We emphasize that a breakdown of higher than second--order
solutions at late non--linear stages
does not contradict the result
that third--order clearly improves upon second--order in the weakly
non--linear regime before shell--crossing, e.g. in
comparison with the exact spherically symmetric solution, or
the skewness and kurtosis moments of the density field (Munshi \etal
1994); here we tested the extreme situation of going to stages
which involve a large number of shell--crossings.

We finally remark that
that we were using the exact expressions for second-- and third--order
solutions for initial conditions, where the peculiar--velocity
potential ${\cal S}$ and the peculiar--gravitational potential $\phi$
are initially proportional (${\cal S} = - \phi t_i$).
The expressions used by other
authors on higher--order Lagrangian schemes refer to different
initial conditions and do not involve terms other than the leading
term. The solutions we use involve
growing parts (even with different signs
than the leading terms) due to second-- and third--order homogeneous
solutions of the basic differential equations.
The leading terms are the particular solutions of these equations only.
Of course, at late non--linear stages analyzed here,
only the leading terms matter, which we confirmed by dropping all
terms other than the leading terms. We found that for expansion factors
$a < 25$ differences with the calculations presented here
become significant (compare, e.g., the second--order coefficient
of the particular solution of the displacement field $-3/14 a^2$
with the growing homogeneous solution $+ 3/5 a$; at $a \approx 25$
the difference still is a $10$\% effect in the time--coefficients).

\vfill\eject

\noindent{\rmm 5. Conclusions}
\bigskip\medskip
\noindent
We have analyzed a family of hierarchical models
with different amount
of power on small scales, and evolved deeply into the non--linear
regime of structure formation. We have compared the
results from N--body simulations with those from Lagrangian perturbation
schemes, where for the latter we have filtered the initial data before
evolving them. The filter scale was determined
by optimization of the
cross--correlation coefficient with the corresponding N--body
simulation.
\medskip
We can report two striking advantages of going to second order in
perturbation theory. Firstly, especially the statistics which probe
the gravitational dynamics of the models show improvement due to
second--order corrections. This success is found for a considerably
higher non--linearity than is expected from a perturbation approach.
Secondly, the improvement (although small for much
small--scale power) is {\it robust} by going to later stages and
to smaller scales.
\medskip
We conclude that the second--order scheme with truncated power--spectra
will be especially useful for modelling
large--scale structure. In particular,
the second--order scheme is computationally simple to realize with
a CPU speed comparable to that of TZA.
Since the second--order corrections to TZA provide noticeable
improvement
of dynamical accuracy for initial data
with negative sloped power--spectra, we expect that
the truncated second--order scheme will be useful for the
modelling
of standard cosmogonies (like Hot--, Cold--, and Mixed--Dark--Matter).
This modelling will be especially effective for large sample
calculations, since in numerical realizations
of `fair' samples in excess of $300$\hn Mpc, performed with
the same resolution as the simulations reported here, the truncation
scale is close to the Nyquist frequency of the N--body
computing. Thus, shortcomings of the analytical schemes become
negligible which puts them in an ideal position for the purpose of
simulating the environment of galaxy formation down to scales,
where other physical effects start to affect models which are based
on the description of self--gravity alone. Our method can be effective
down to
galaxy group mass scales ($10^{13} M_\odot$), or better if we include
biasing or
go to epochs earlier than the present.

\smallskip\bigskip\noindent
{\rma Acknowledgements:} {\mfst We would like to thank Robert Klaffl
for valuable discussions on numerical problems.

\noindent
TB is supported by DFG (Deutsche Forschungsgemeinschaft). ALM wishes to
acknowledge support from NASA and from NSF grants AST--9021414 and
OSR--9255223, and facilities of the National Center for Supercomputing
Applications, all in the USA.
We are grateful to the Aspen Center for Physics
(USA) for its June 1994 workshop on topics related to this work.}

\vfill\eject

\def\ref{\par\noindent\hangindent\parindent\hangafter1}
\centerline{\rmm References}
\bigskip\bigskip
{\mfst
\ref
Bagla J.S., Padmanabhan T. (1994): {\it M.N.R.A.S.} {\bf 266}, 227.
\smallskip
\ref
Bernardeau F. (1994): {\it Ap.J.}, in press.
\smallskip
\ref
Bertschinger E., Jain B. (1994): {\it Ap.J.}, in press.
\smallskip
\ref
Bertschinger E., Hamilton A.J.S. (1994): {\it Ap.J.}, submitted.
\smallskip
\ref
Brainerd T.G., Scherrer R.J., Villumsen J.V. (1994) {\it Ap.J.}
{\bf 418}, 570.
\smallskip
\ref
Bouchet F.R., Juszkiewicz R., Colombi S., Pellat R. (1992):
{\it Ap.J. Lett.} {\bf 394}, L5.
\smallskip
\ref
Buchert T. (1989): {\it Astron. Astrophys.} {\bf 223}, 9.
\smallskip
\ref
Buchert T. (1992): {\it M.N.R.A.S.} {\bf 254}, 729.
\smallskip
\ref
Buchert T. (1993a): {\it Astron. Astrophys. Lett.} {\bf 267}, L51.
\smallskip
\ref
Buchert T. (1993b): in: {\it 4th. MPG--CAS workshop on High--energy
Astrophysics and Cosmology}, eds.: G. B\"orner, T. Buchert, MPA/P8,
pp. 204--214.
\smallskip
\ref
Buchert T. (1994): {\it M.N.R.A.S.}, in press.
\smallskip
\ref
Buchert T., Ehlers J. (1993): {\it M.N.R.A.S.} {\bf 264}, 375.
\smallskip
\ref
Buchert T., Melott A.L., Wei{\ss} A.G. (1994a): {\it Astron.
Astrophys.}, in press; (BMW).
\smallskip
\ref
Buchert T., Karakatsanis G., Klaffl R., Schiller P. (1994b):
in preparation.
\smallskip
\ref
Coles P., Melott A.L., Shandarin S.F. (1993): {\it M.N.R.A.S.}
{\bf 260}, 765.
\smallskip
\ref
Croudace K.M., Parry J., Salopek D.S., Stewart J.M. (1994): {\it
Ap.J.}, in press.
\smallskip
\ref
Ehlers J., Buchert T. (1994): {\it M.N.R.A.S.}, to be submitted.
\smallskip
\ref
Giavalisco M., Mancinelli B., Mancinelli P.J., Yahil A. (1993):
{\it Ap.J.} {\bf 411}, 9.
\smallskip
\ref
Gramann M. (1993): {\it Ap.J. Lett.} {\bf 405}, 47.
\smallskip
\ref
Kasai M. (1993): {\it Phys. Rev.} {\bf D47}, 3214.
\smallskip
\ref
Kofman L.A., Pogosyan D., Shandarin S.F., Melott A.L. (1992):
{\it Ap.J.} {\bf 393}, 437.
\smallskip
\ref
Kofman L.A., Pogosyan D. (1994): {\it Ap.J.}, submitted.
\smallskip
\ref
Lachi\`eze-Rey M. (1993a): {\it Ap.J.} {\bf 407}, 1.
\smallskip
\ref
Lachi\`eze-Rey M. (1993b): {\it Ap.J.} {\bf 408}, 403.
\smallskip
\ref
Matarrese S., Lucchin F., Moscardini L., Saez D. (1992): {\it
M.N.R.A.S.} {\bf 259}, 437.
\smallskip
\ref
Matarrese S., Pantano O., Saez D. (1993): {\it Phys. Rev.}
{\bf D47}, 1311.
\smallskip
\ref
Matarrese S., Pantano O., Saez D. (1994): {\it M.N.R.A.S.}, in press.
\smallskip
\ref
Melott A.L. (1986): {\it Phys. Rev. Lett.} {\bf 56}, 1992.
\smallskip
\ref
Melott A.L. (1993): {\it Ap.J. Lett.} {\bf 414}, L73.
\smallskip
\ref
Melott A.L., Pellman T.F., Shandarin S.F. (1994a): {\it M.N.R.A.S.},
in press; (MPS).
\smallskip
\ref
Melott A.L., Lucchin F., Matarrese S., Moscardini L. (1994b)
{\it M.N.R.A.S.}, in press.
\smallskip
\ref
Melott A.L., Shandarin S.F., Weinberg D.H. (1994c): {\it Ap.J.},
in press.
\smallskip
\ref
Moutarde F., Alimi J.-M., Bouchet F.R., Pellat R., Ramani A.
(1991): {\it Ap.J.} {\bf 382}, 377.
\smallskip
\ref
Munshi D., Starobinsky A.A. (1994): {\it Ap.J.}, in press.
\smallskip
\ref
Munshi D., Sahni V., Starobinsky A.A. (1994): {\it Ap.J.}, submitted.
\smallskip
\ref
Park C. (1990): {\it Ph.D. Thesis}, Princeton.
\smallskip
\ref
Park C. (1991): {\it M.N.R.A.S.} {\bf 251}, 167.
\smallskip
\ref
Peebles P.J.E. (1980): {\it The Large-scale Structure of the Universe},
Princeton Univ. Press.
\smallskip
\ref
Salopek D.S., Stewart J.M., Croudace K.M. (1994): {\it M.N.R.A.S.},
submitted.
\smallskip
\ref
Shandarin S.F., Zel'dovich Ya.B. (1989): {\it Rev. Mod. Phys.} {\bf
61}, 185.
\smallskip
\ref
Weinberg D.H., \etal (1994): work in progress.
\smallskip
\ref
Zel'dovich Ya.B. (1970): {\it Astron. Astrophys.} {\bf 5}, 84.
\smallskip
\ref
Zel'dovich Ya.B. (1973): {\it Astrophysics} {\bf 6}, 164.

}

\vfill\eject

\centerline{\rmm Figure Captions}
\bigskip\medskip
{\mfst
\noindent{\bf Figure 1:} The variation of the cross--correlation
coefficient $S$ (eq. (5)) against the truncation scale $k_c$ imposed on
the initial data for the evolved analytical schemes ($k_{nl}=8k_f$)
for different
filter types (sharp $k-$truncation (kc; eq. (6a)), Gaussian
truncation (gs; eq. (6b)), and (real space)
top--hat truncation (th; eq. (6c))), for spectral indices
$n=+1,0,-1,-2,-3$ (Fig.1a,b,c,d,e).
The first--order model is shown as a dotted line, the second--order
model as a dashed line.

\bigskip

\noindent{\bf Table 1:} The cross--correlation coefficient $S$ for
the optimal truncation scale $k_{opt}$ in units of $k_{nl}=8k_f$
(stage f), and $k_{nl}=4k_f$ (stage g)
for different spectral indices
and the two Lagrangian perturbation schemes.

\bigskip

\noindent{\bf Figure 2:} Illustration to Table 1.
Shown is the
cross--correlation coefficient $S$ for the optimal truncation scale
$k_{opt}$ according to Table 1 for stage f (left) and stage g (right)
as a function of spectral index.
The first--order scheme is shown with squares,
the second--order scheme with diamonds.

\bigskip

\noindent{\bf Figure 3:} The cross--correlation coefficient $S$ for
the stage $k_{nl}=8k_f$ and
the cases $n=-1$ (upper) and $n=+1$ (lower) as a function of
scale. Displayed is the cross--correlation of the N--body simulation
with the
Eulerian linear extrapolation of the initial data (dashed--dotted),
the ``chopped'' Eulerian linear approximation
(thick dashed--dotted), the
untruncated first-- (dotted) and second--order (dashed)
approximations, and the optimally truncated first-- (thick dotted) and
second--order (thick dashed) approximations.

\bigskip

\noindent{\bf Figure 4:} A thin slice (thickness $L/64$) of the
density field is shown for the numerical (upper--left), the
``chopped'' Eulerian linear (upper--right), the
optimally truncated first--order
(middle--left) and second--order (lower--left), the
untruncated first--order (middle--right), and the
untruncated second--order (lower--right)
approximations for
the evolution stage corresponding to $k_{nl}=8 k_f$, and for
different power--spectra
with index $n=+1,0,-1,-2,-3$ (Figs. 4a,b,c,d,e).
The grey--scale is linear, and the
maximum density contrasts are adapted to give a satisfactory visual
impression, except for the optimally truncated models for which
the upper truncation
is the same (density contrast $5$)
as for the numerical simulation to allow for an
objective comparison.

\bigskip

\noindent{\bf Figure 5:} The standard deviations of the density
contrast as a function of smoothing scale (given in grid units
$R_G$ of a $64^3$ grid)
in the optimally truncated first-order (dotted), the optimally
truncated second--order (dashed), and the
numerical (full line) approximations, for the different power--spectra
$n=-3,-2,-1,0,+1$ (Figs. 5a,b,c,d,e,).

\bigskip

\noindent{\bf Figure 6:} The cross--correlation coefficient $S$
as a function of the standard deviation $\sigma_{\rho}$ of the
numerical simulation for
the different power--spectra $n=-3,-2,-1,0,+1$ (Figs. 6a,b,c,d,e). The
cross--correlation of the N-body
with the optimally truncated first--order model is
shown as a dotted line; with the optimally truncated second--order
model a dashed line.

\bigskip

\noindent{\bf Figure 7:} The power--spectra of the N--body simulations
(solid line) compared with the optimally truncated first-- (dotted) and
second--order (dashed) approximation schemes for the
different power--spectra $n=-3,-2,-1,0,+1$ (Figs. 7a,b,c,d,e).

\bigskip

\noindent{\bf Figure 8:} The relative phase--errors. Notation and
labelling like in Figure 6.

\bigskip

\noindent{\bf Figure 9:} The density distribution functions.
Notation and labelling like in Figure 7.

\bigskip

\noindent{\bf Figure 10:} Same as Fig.5 for a later stage when the
non--linearity scale has dropped to half the value of the former stage
($k_{nl} = 4k_f$).

\bigskip

\noindent{\bf Figure 11:} Same as Fig.6 for the later stage.

\bigskip

\noindent{\bf Figure 12:} Same as Fig.7 for the later stage.

\bigskip

\noindent{\bf Figure 13:} Same as Fig.8 for the later stage.

\bigskip

\noindent{\bf Figure 14:} Same as Fig.9 for the later stage.

}

\vfill\eject
\bye